\documentclass[11pt]{article}
\usepackage[utf8]{inputenc}

\usepackage{amssymb,siunitx,amsmath,bm}
\usepackage[left=2cm,right=2cm,top=2cm,bottom=2cm]{geometry}
\usepackage{graphicx}

\usepackage{tikz}
\usetikzlibrary{shapes,arrows}

\usepackage{natbib}
\setcitestyle{authoryear}

\usepackage[symbol]{footmisc}

\newcommand{\pdev}[2]{\frac{\partial #1}{\partial #2}}

\newcommand\Rey{\mbox{\textit{Re}}}  
\newcommand\Ri{\mbox{\textit{Ri}}}   
\newcommand\Prn{\mbox{\textit{Pr}}}   
\newcommand\Scn{\mbox{\textit{Sc}}}
\newcommand\Nus{\mbox{\textit{Nu}}}

\definecolor{col1t}{rgb}{0,0,1}
\colorlet{col1}{white!70!col1t}
\definecolor{col2t}{rgb}{1,0,1}
\colorlet{col2}{white!80!col2t}
\definecolor{col3t}{rgb}{1,0,0}
\colorlet{col3}{white!60!col3t}

\begin{document}

{
\vspace{2cm}
\centering

\Large
\textbf{Transport by waves and turbulence: Dilute suspensions in stably stratified plane Poiseuille flow} \\
\vspace{0.5cm}
}
{
\centering
\textbf{Charlie Lloyd\textsuperscript{1}\footnote[2]{Email for correspondence: C.J.Lloyd@hull.ac.uk} and Robert Dorrell\textsuperscript{2}} \\ 
}
\vspace{0.5cm}
{
\small
\noindent
\textsuperscript{1} Energy and Environment Institute, University of Hull, Hull, HU6 7RX \\
\textsuperscript{2} School of Architecture, Building and Civil Engineering, Loughborough University, Loughborough, LE11 3TU \\
\hrule
}

\section*{Abstract}
The transport of particulate material is fundamental to a wide range of industrial and environmental flows, from energy generation to manufacturing and from marine to atmospheric systems. All such flows are inherently density stratified where varying concentration of particulates changes the density of the fluid; however, the role of density stratification on particle transport remains poorly understood.
Here we investigate the transport dynamics of negatively buoyant passive sediment in a stably stratified plane Poiseuille flow using numerical simulation.
The sediment is passive, making no contribution to flow density; however, the flow is thermally stratified. This set up is chosen to quantify how stratification controls sediment transport processes.
Sediment concentration fields are passively solved with settling velocities of $V_s = 0.005$, $V_s = 0.01$ and $V_s = 0.02$, made dimensionless by the friction velocity $u_\tau$. 
Strong buoyancy forces in the core have a profound impact on sediment transport, leading to two-layer sediment concentration profiles with concentration gradients increasing with sediment settling velocity, $V_s$. 
When compared to unstratified flow, stratification leads to considerably larger differences in concentration profiles and higher order statistics between the three different sediments. 
However, when scalar statistics are appropriately scaled by either $\theta_\tau$ for temperature or $V_s \overline{c}$ for sediment, where $\theta_\tau$ is the thermal shear temperature and $\overline{c}$ is the vertically varying mean sediment concentration, all scalars collapse to near common vertically varying profiles. 
Collapse is explained by revisiting the gradient diffusion hypothesis, which links vertical scalar fluxes to their mean vertical scalar gradients.
However, collapse of scalars is poor in the core of the channel, where large concentration gradients coincide with large-scale mixing events.
Here the differences between vertical turbulent diffusivities of sediment and temperature increase with increasing $V_s$, reaching \SI{20}{\%} for $V_s = 0.02$.
Discrepancies arise due to a breakdown of the linear gradient diffusion hypothesis.
Predictions using the gradient diffusion hypothesis are expected to worsen with increasing settling velocity, indicating that classical models poorly predict dilute particle transport in strongly stratified flows, which are common in a range of environmental and industrial settings.

\section{Introduction}
\label{section:introduction}
An improved understanding of particle transport has applications spanning engineering, atmospheric, oceanographic and earth surface sciences.
improved models for particulate transport are critical for our understanding of microplastic transport in the atmosphere and marine/fresh-water ecosystems \citep{van2015global,evangeliou2020atmospheric}, sediment and tracer (e.g. organic matter) transport in rivers and oceans \citep{wu2000nonuniform,morrison2022ventilation}, turbid underwater currents \citep{dorrell2019self}, and atmospheric currents in the form of powder snow avalanches \citep{bartelt2016configurational} and pyroclastic flows \citep{sulpizio2014pyroclastic}.
These flows are highly complex and comprised of a wide range of turbulent length and time-scales, and are further complicated by high volume fraction loading, where particles may interact and have a primary control on fluid dynamics. 
Even in dilute flows density stratification is commonplace, which can have a strong influence on scalar transport \citep{dorrell2019self}. 
Stable stratification may arise through a compositional change in flow properties (e.g. salinity, temperature) or through suspended particulates. 
These can lead to regions of strong buoyancy gradient or interfaces which introduce strong anisotropy due to suppressed vertical motion, and subsequently barriers to vertical transport. 
In addition, it is common for strong buoyancy gradients to introduce large-scale coherent structures in the form of waves and vortices \citep{buhler2014waves}. 
It has been theorised that large-scale coherent structures play a critical role in sediment transport processes, and may be a ubiquitous feature of many fluid dynamical flows \citep{caulfield2021layering}, yet the role of coherent structures on sediment transport is poorly understood \citep{wells2021turbulence}. 

Of particular importance for sediment transport predictions is the quantification and modelling of sediment diffusivity due to turbulence, $K_c$, derived from the gradient diffusion hypothesis \citep{rouse1937modern}. 
Our ability to predict sediment transport dynamics at large scales is critically dependent on appropriate parameterisation of $K_c$, with simple closures based upon the canonical log-law \citep{rouse1937modern}, and more advanced models developed for stratified flows \citep{mclean1992calculation}.
While these closures are commonly adopted in sediment transport models, it has been noted that they may be inappropriate in regions where density, as determined by sediment concentration, changes rapidly when compared to the scale of mixing events \citep{nielsen2004turbulent}.
The gradient diffusion hypothesis assumes that mixing length scales are small compared to the length scales over which concentration varies, and situations may therefore arise when such conditions do not hold.

This study investigates the role of large-scale coherent structures on sediment transport by simulating a weakly stratified plane Poiseuille (or channel) flow.
The plane Poiseuille flow is doubly periodic flow driven by a streamwise pressure gradient between two walls; a canonical fluid dynamics problem that has received considerable interest for studying wall-bounded turbulence
\citep{kim1987turbulence,moser1999direct,del2003spectra,vreman2014comparison,jimenez2022streaks}, and to investigate sediment and scalar transport processes in wall bounded flows \citep{cantero2009turbidity,cantero2009direct,dutta2014effect}.
In the context of sediment transport processes, stable stratification is typically associated with a downward flux of dense settling particles which have a concentration high enough to influence the flow through a buoyancy term in the momentum equations (i.e the system is one-way coupled). 
In this case, asymmetry may develop in the flow as the sediment concentration gradients suppress vertical motion; the position of maximum flow velocity moves downwards as a response to reduced turbulence, anisotropy of the Reynolds stresses increases, vertical transport of sediment is suppressed therefore increasing associated concentration gradients, and a turbulent flow may locally re-laminarise if the sediment settling velocity and concentration are sufficiently large \citep{cantero2009direct,dutta2014effect}.

Here a channel flow where stratification is driven by a temperature field is investigated, with fixed temperature boundary conditions at the upper and lower walls.
Buoyancy forces are dependent on the flow temperature, and the stratification strength is therefore governed by the temperature difference between the upper and lower walls.
Subsequently, the flow is parameterised by the friction Reynolds number, the friction Richardson number, and the Prandtl number, respectively defined as
\begin{equation}
    \Rey_\tau = \frac{u_\tau \delta}{\nu}, \ \ \ \Ri_\tau = \frac{\alpha \Delta T g \delta}{u_\tau^2}, \ \ \ \Pr = \frac{\kappa_\theta}{\nu},
\end{equation}
where $u_\tau$ is the friction velocity, $\delta$ the channel half-height, $\nu$ represents kinematic viscosity, $\Delta T$ is the temperature difference between the upper and lower walls, $g$ represents gravitational acceleration, $\alpha$ is the thermal expansion coefficient, and $\kappa_\theta$ represents thermal diffusivity. 
Under appropriate Reynolds and Richardson numbers, the stratified channel flow possesses a range of coherent structures due to the large buoyancy forces that dominate the core of the channel, and strong shear at the walls.
This flow introduces large-scale internal waves in the channel core, driven by hairpin vortex ejections in the outer regions \citep{lloyd2022coupled,lloyd2024linear}, which are emerging as a common feature in stratified shear flows 
\citep{watanabe2019hairpin,salinas2021anatomy}
.
The coexistence of relatively large-scale coherent structures and strong buoyancy gradients makes the thermally stratified channel flow an interesting test-case for investigating sediment transport processes \citep{boetti2023pair}. 

This study quantifies passively transported, negatively (or positively, due to symmetry) buoyant sediment in a thermally stratified channel flow, numerically simulated using spectral element methods.
Through this study we develop an understanding of sediment transport processes in stratified shear flows comprising both active turbulence and large-scale coherent structures, free from complexities associated with a fully coupled, non-dilute, sediment concentration field, which we reserve for future work.
It is shown that while stratification has a significant effect on sediment transport, introducing a strong transport barrier in the channel core, turbulent statistics collapse to near common profiles when appropriately scaled by $\theta_\tau$ for temperature and $V_s \overline{c}$ for sediment, where $V_s$ is the particle settling velocity and $\overline{c}$ is the temporally and planar-averaged vertically varying sediment concentration.
Deviations in statistics only appear in the channel core, where buoyancy forces are strong and sediment concentration gradients are large.
We argue that deviations arise due to a breakdown of the gradient diffusion hypothesis, where higher order terms containing $\partial_y^3 \overline{c}$ become important, or alternatively, the collapse of scalar statistics requires $\overline{c'v'} \sim \partial_y \overline{c}$ to hold, which is only weakly true in the core of the channel.

This paper is organised as follows: The methodology is introduced in Section \ref{section:methodology}, before reporting our results in Section \ref{section:results}. Vertically varying turbulent statistics are presented in Section \ref{section:timeStats} before discussion multi-dimensional spectra in Section \ref{section:spectra} and Spectral Proper Orthogonal Decompositions in Section \ref{section:SPOD}.
Finally we present visualisations of instantaneous flow in Section \ref{section:inst}, before concluding our study in Section \ref{section:conclusions}.

\begin{figure}
\centering
\def\constone{-5.3}
\begin{tikzpicture}
\draw [fill=col1,draw=none] (0,-1.5) rectangle (13,-1.2);
\draw [fill=col2,draw=none] (0,-1.2) rectangle (13,-0.3);
\draw [fill=col3,draw=none] (0,-0.3) rectangle (13,0.3);
\draw [fill=col2,draw=none] (0,0.3) rectangle (13,1.2);
\draw [fill=col1,draw=none] (0,1.2) rectangle (13,1.5);

\draw[black,very thick] (0,-1.5) -- (13,-1.5);
\draw[black,very thick] (0,1.5) -- (13,1.5);
\draw[black,thick,dashed] (0,0) -- (13,0);

\draw[thick,->] (0.5,0) -- (1.5,0) node[anchor= north east] {$x$};
\draw[thick,->] (0.5,0) -- (0.5,1) node[anchor= north east] {$y$};

\node at (1,-1.8){$y=0$};
\node at (1,1.8) {$y=2$};
\node at (10.5,1.8) {$Sc^{-1} Re_\tau^{-1} \partial_y c = - V_s c$};
\node at (10.5,-1.8) {$Sc^{-1} Re_\tau^{-1} \partial_y c = - V_s c$};
\node at (7,-1.8){$\theta=-\frac{1}{2}$};
\node at (7,1.8) {$\theta=\frac{1}{2}$};
\node at (4.0,-1.8){$U=V=W=0$};
\node at (4.0,1.8) {$U=V=W=0$};

\draw plot [smooth] coordinates {
(3.0, -1.5) (3.539, -1.44) (3.652, -1.38) (3.707, -1.32) (3.748, -1.26) (3.782, -1.2) (3.814, -1.14) (3.845, -1.08) (3.874, -1.02) (3.903, -0.96) (3.932, -0.9) (3.962, -0.84) (3.992, -0.78) (4.023, -0.72) (4.056, -0.66) (4.091, -0.6) (4.129, -0.54) (4.171, -0.48) (4.218, -0.42) (4.272, -0.36) (4.333, -0.3) (4.403, -0.24) (4.479, -0.18) (4.552, -0.12) (4.602, -0.06) (4.619, 0.0) (4.602, 0.06) (4.551, 0.12) (4.479, 0.18) (4.402, 0.24) (4.332, 0.3) (4.271, 0.36) (4.218, 0.42) (4.171, 0.48) (4.129, 0.54) (4.091, 0.6) (4.055, 0.66) (4.023, 0.72) (3.991, 0.78) (3.961, 0.84) (3.932, 0.9) (3.903, 0.96) (3.874, 1.02) (3.845, 1.08) (3.814, 1.14) (3.782, 1.2) (3.747, 1.26) (3.707, 1.32) (3.651, 1.38) (3.539, 1.44) (3.0, 1.5) 
};

\draw plot [smooth] coordinates {
(6.0, -1.5) (6.098, -1.44) (6.12, -1.38) (6.131, -1.32) (6.139, -1.26) (6.146, -1.2) (6.152, -1.14) (6.159, -1.08) (6.165, -1.02) (6.173, -0.96) (6.18, -0.9) (6.189, -0.84) (6.198, -0.78) (6.209, -0.72) (6.222, -0.66) (6.237, -0.6) (6.255, -0.54) (6.278, -0.48) (6.308, -0.42) (6.345, -0.36) (6.395, -0.3) (6.464, -0.24) (6.563, -0.18) (6.694, -0.12) (6.845, -0.06) (7.001, 0.0) (7.157, 0.06) (7.308, 0.12) (7.438, 0.18) (7.536, 0.24) (7.605, 0.3) (7.655, 0.36) (7.693, 0.42) (7.722, 0.48) (7.745, 0.54) (7.763, 0.6) (7.778, 0.66) (7.791, 0.72) (7.802, 0.78) (7.811, 0.84) (7.82, 0.9) (7.828, 0.96) (7.835, 1.02) (7.841, 1.08) (7.848, 1.14) (7.854, 1.2) (7.861, 1.26) (7.869, 1.32) (7.88, 1.38) (7.902, 1.44) (8.0, 1.5) 
};

\draw plot [smooth] coordinates {
(10.89, -1.5) (10.45, -1.44) (10.366, -1.38) (10.325, -1.32) (10.296, -1.26) (10.272, -1.2) (10.249, -1.14) (10.227, -1.08) (10.205, -1.02) (10.182, -0.96) (10.157, -0.9) (10.131, -0.84) (10.102, -0.78) (10.07, -0.72) (10.033, -0.66) (9.992, -0.6) (9.943, -0.54) (9.887, -0.48) (9.82, -0.42) (9.741, -0.36) (9.649, -0.3) (9.54, -0.24) (9.417, -0.18) (9.293, -0.12) (9.193, -0.06) (9.126, 0.0) (9.082, 0.06) (9.055, 0.12) (9.038, 0.18) (9.029, 0.24) (9.024, 0.3) (9.021, 0.36) (9.019, 0.42) (9.017, 0.48) (9.016, 0.54) (9.016, 0.6) (9.015, 0.66) (9.015, 0.72) (9.014, 0.78) (9.014, 0.84) (9.013, 0.9) (9.013, 0.96) (9.013, 1.02) (9.013, 1.08) (9.012, 1.14) (9.012, 1.2) (9.012, 1.26) (9.012, 1.32) (9.011, 1.38) (9.011, 1.44) (9.008, 1.5) 
};

\node at (9.5,0.75) {$c(y)$};

\node at (7.0,0.75) {$\theta(y)$};

\draw[thick,->] (8.2 + \constone, 0.6) -- (9 + \constone, 0.6);
\draw[thick,->] (8.2 + \constone, 0.3) -- (9 + \constone, 0.3);
\draw[thick,->] (8.2 + \constone, 0.0) -- (9 + \constone, 0.0);
\draw[thick,->] (8.2 + \constone,-0.3) -- (9 + \constone,-0.3);
\draw[thick,->] (8.2 + \constone,-0.6) -- (9 + \constone,-0.6);
\node at (4.6,0.75) {$U(y)$};
\end{tikzpicture}
\caption{Numerical domain and simulation boundary conditions. Shaded regions represent the inner regions at $y\leq0.2$ and $y\geq 1.8$, the outer regions at $0.2 < y \leq 0.8$ and $1.2 \leq y < 1.8$, and the channel core at $0.8 < y < 1.2$. The dashed line represents the channel centreline at $y=1$.
}
\label{figure:geom}
\end{figure}
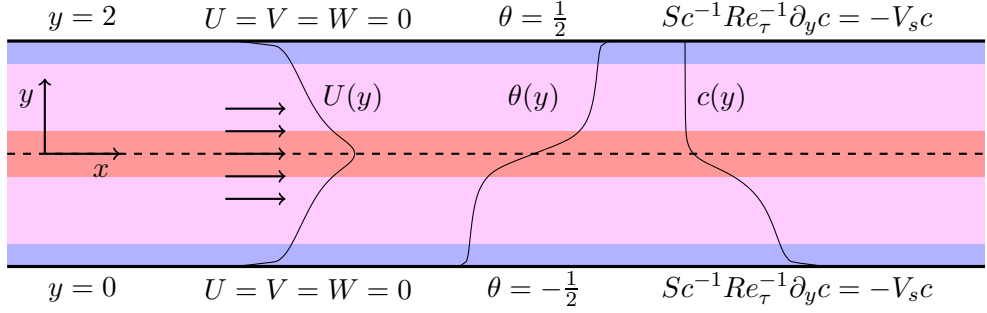

\section{Methodology}
\label{section:methodology}
A stratified channel flow is simulated (Figure \ref{figure:geom}) driven by a constant streamwise (negative) pressure gradient with periodic boundary conditions applied in the streamwise ($x$) and spanwise ($z$) directions and no-slip conditions at the walls ($y-$normal boundaries). 
Stratification is imposed with a fixed temperature ($T$) difference between the upper and lower walls which affects the density field, $\rho$, by
\begin{equation}
    \rho = \rho_0 -\alpha \rho_0 (T - T_0),
\end{equation}
where $\rho_0$ and $T_0$ represent reference values for density and temperature, respectively, and $\alpha$ is the thermal expansion coefficient.
We assume the flow is incompressible and Boussinseq, such that the governing dimensionless equations are the continuity, momentum, and temperature scalar transport equations:
\begin{equation}
\label{eq:mom}
    \frac{\partial U_i}{\partial t} + U_j \frac{\partial U_i}{x_j} = -\frac{\partial P}{\partial x_i} + \frac{1}{\Rey_\tau} \frac{\partial^2 U_i}{\partial x_j^2} + f_i + \Ri_\tau \theta' \hat{y}_i.
\end{equation}
\begin{equation}
    \frac{\partial U_i}{\partial x_j} = 0,
\end{equation}
\begin{equation}
\label{eq:temp}
    \frac{\partial \theta}{\partial t} + U_j \frac{\partial \theta}{\partial x_j} = \frac{1}{\Prn \Rey_\tau}\frac{\partial^2 \theta}{\partial x_j^2}.
\end{equation}
where  $U_i=(U,V,W)'$  represents velocity,  $x_i=(x,y,z)'$  represents Cartesian spatial coordinates,  $t$  represents time,  $P$  represents kinematic pressure,  $\theta = (T-T_0)/\Delta T$ is the temperature field, and  $\hat{y}_i$  is the  $y$-direction unit vector. The flow is forced by a constant streamwise (negative) pressure gradient,  $f_i=(1,0,0)'$. 
Equations are made dimensionless by the channel half height, $\delta$, the friction velocity $u_\tau$, and the temperature difference between the upper and lower walls, $\Delta T$, which is fixed with dirichlet boundary conditions ($\theta = -0.5$ at $y=0$ and $\theta = 0.5$ at $y=2$). 
Walls are treated as no-slip with $U_i = 0$.
The dimensionless parameters governing dynamics are the shear Reynolds number $\Rey_\tau$, the shear Richardson number, $\Ri_\tau$, and the Prandtl number, $\Prn$:
\begin{equation}
    \Rey_\tau = \frac{u_\tau \delta}{\nu}, \ \ \ \Ri_\tau = \frac{\alpha \Delta T g \delta}{u_\tau^2}, \ \ \ \Pr = \frac{\kappa_\theta}{\nu},
\end{equation}
where $g$ represents gravitational acceleration, $\kappa_\theta$ represents thermal diffusivity, and $\nu$ represents momentum diffusivity (viscosity). 
Following \citet{lloyd2022coupled}, simulations are performed with fixed Reynolds and Prandtl number, $\Rey_\tau = 550$ and $\Prn = 1$, and two Richardson numbers, $\Ri_\tau = 0$ and $\Ri_\tau = 480$, corresponding to an unstratified channel flow with a passive temperature field, and a stratified channel flow with shear-dominated regions near the walls and a buoyancy dominated channel core. 

The buoyancy force $\Ri_\tau \theta' \hat{y}_i$ is dependent on the temperature fluctuation, $\theta' = \theta - \overline{\theta}$, where $\overline{\theta}$ is the time and planar-averaged temperature field. In this case the purely wall-normal temperature field $\overline{\theta}$ is absorbed hydrostatically into the pressure term. 
In practice $\overline{\theta}$ is approximated at each timestep by the instantaneous planar-averaged temperature, consistent with previous work \citep{armenio2002investigation,garcia2011turbulence,lloyd2022coupled}.
Equations \eqref{eq:mom} to \eqref{eq:temp} govern system dynamics, and are analogous to the system of equations solved by \citet{lloyd2022coupled}, aside from our choice of passive scalar, where we choose to solve for a temperature field rather than a density field.

Dilute sediment is transported passively in this system by solving additional transport equations of the form
\begin{equation}
\label{eq:sed_scalar}
    \frac{\partial c}{\partial t} + U_j \frac{\partial c}{\partial x_j} = \frac{1}{\Scn \Rey_\tau}\frac{\partial^2 c}{\partial x_j^2} + V_s \frac{\partial c}{\partial y},
\end{equation}
where $c$ is the sediment concentration (made dimensionless by the volume averaged concentration $c^v$), $\Scn = \kappa_c/\nu$ is the Schmidt number where $\kappa_c$ is the non-turbulent sediment diffusivity, and $V_s$ is the sediment settling velocity.
The sediment concentration scalar transport equation \eqref{eq:sed_scalar} is obtained by assuming i) particles are smaller than the smallest turbulent length scales in the flow, ii) the aerodynamics response time of the particles is small compared to the smallest time scales of the flow, iii) sediment concentrations are dilute such that particle-particle interactions and the effects of particle suspension on buoyancy can be neglected, and iv) the particles are monodisperse. 
Under these assumptions the velocity field remains divergence-free, fluid density/buoyancy is only affected by the temperature field, and particles settle at a constant velocity, $V_s$, determined by the balance between the gravitational force and the Stokes drag force acting on individual particles.
The advection term of \eqref{eq:sed_scalar} captures the mean flow transport and, crucially, the turbulent diffusion of particles across scales from dissipation of energy to flow scale. 
The diffusion term of \eqref{eq:sed_scalar} captures the decay of concentration gradients due to hydrodynamic diffusion of particles, Brownian motion, and variations in particle inertia (size/shape) \citep{meiburg2015modeling,ham1988hindered,davis1988spreading}. 
Here we specify $\Scn = \Prn = 1$ for all simulations and scalars, following previous work where the influence of Schmidt number was found to be negligible so long as $\Scn \geq 1$ \citep{necker2005mixing, nasr2011turbins,nasr2014turbidity, pelmard2018grid}.
For a more complete discussion on these assumptions please see the review of \citet{meiburg2015modeling} and study of \citet{marshall2021effect}. 

Boundary conditions are periodic for all variables in $x$ and $z$. At the walls velocity boundary conditions are no-slip ($U_i = 0$), and the temperature field is $\theta = -0.5$ at $y=0$ and $\theta = 0.5$ at $y=2$, creating a stably stratified shear flow.
Sediment concentration boundary conditions are zero-flux to ensure the volume averaged concentration field is constant in time. At the walls,
\begin{equation}
    \frac{1}{\Scn \Rey_\tau}\frac{\partial c}{\partial y} = -V_s c.
\end{equation}

Equations are discretised and solved using Nek5000, following the methodology of \citet{lloyd2022coupled}.
The domain size is $L_x \times L_y \times L_z = 8\pi \times 2 \times 3\pi$ (normalised by the channel half height), decomposed into $N_x \times N_y \times N_z = 80 \times 44 \times 40$ spectral elements, stretched vertically with a Chebyshev distribution, and further discretised by $8^3$ Gauss-Lobatto-Legendre (GLL) nodes (seventh-order polynomials).
Third-order time-integration is adopted with a dimensionless timestep of $10^{-4}$. 
This choice of grid resolution follows \citet{lloyd2022coupled}; while vertical resolution is high enough to fully resolve turbulent scales the spanwise and streamwise directions are under-resolved. Sub-grid-scale dissipation is accounted for using explicit modal based filtering.
While this choice of resolution is inadequate for resolving the inner regions of the flow where shear is largest, \citet{lloyd2022coupled} obtained excellent agreement with DNS in the outer and core-regions of the channel ($0.2 < y < 1.8$), which are the regions of interest for this study. 
For further details and validation of this modelling strategy, see \citet{lloyd2022coupled} and the references therein. 

Two cases are presented in this work, both with $\Rey_\tau = 550$, and three sediment scalars with $V_s = 0.005, 0.01, 0.02$. An unstratified flow (Case U) is simulated with $\Ri_\tau = 0$, initialised from a perturbed logarithmic velocity profile. A stratified flow (Case S) with $\Ri_\tau = 480$ is initialised from psuedo-steady Case U data. 
Both cases are advanced to a statistically steady state in two stages: Firstly, a coarse simulation is performed, using $6^3$ GLL points per element; and secondly, the simulation is run with the desired polynomial order ($8^3$ GLL points).
Both steps are advanced to psuedo-steady state, assessed by monitoring the friction Reynolds numbers $\Rey_\tau$ and wall sediment concentrations. 
Once pseudo-steady, statistics are collected over time $T$ (Table \ref{table:cases}).
Temporal statistics are obtained by collecting instantaneous velocity, temperature, and sediment concentration on 2D slices every 20 timesteps. Several $y-$normal slices are collected, along with a $z-$normal slice at $z=L_z/2$.
over time $T$, 22000 snapshots are collected for Case U and 37500 snapshots are collected for Case S, enabling convergence of multi-dimensional spectra and spectral proper orthogonal decomposition (SPOD). 

\begin{table}
\centering
\caption{
Cases with time-averaged parameters and case-specific quantities.
The bulk Reynolds and Richardson numbers are given by $\Rey_b = U_b \delta/\nu$ and $\Ri_b = \alpha \Delta T g \delta/2  U_b^2 $ where $U_b$ is the bulk velocity.
$\Nus = 2 \partial_y \overline{\theta}|_w$ is the Nusselt number, and $\Lambda = 2 \Rey_\tau \Prn / \varkappa \Ri_\tau \Nus$ is the Obukhov length normalised by the channel half height $\delta$ with the K\'{a}rm\'{a}n constant $\varkappa = 0.41$.
$\Delta^+_i$ quantifies the element sizes in wall units (normalisation by $\delta_\nu=\nu/u_\tau$).
Note that each element is further discretised by $8^3$ GLL points.
}
\label{table:cases}
\begin{tabular}{c c c c c c c c c c c c c c}
Case & $\Rey_\tau$ & $\Rey_b$ & $\Ri_\tau$ & $\Ri_b$  & $\Nus$ & $\Lambda$ & $N_x$ & $N_y$ & $N_z$ & $\Delta_x^+$ & $\Delta_{y,\text{max}}^+$ & $\Delta_z^+$ & $T$ \\
\hline
U & 550 & 10 507 & 0    & 0     & 21.30     & -       & 80 & 44 & 40 & 172.8 & 40.2 & 129.6 & 44 \\
S & 550 & 14 283 & 480  & 0.356 & 4.13      & 1.35    & 80 & 44 & 40 & 172.8 & 40.2 & 129.6 & 75 \\
\hline
\end{tabular}
\end{table}
\section{Results}
\label{section:results}
\subsection{Time- and planar-averaged statistics}
\label{section:timeStats}
\begin{figure}
    \centering
    \includegraphics{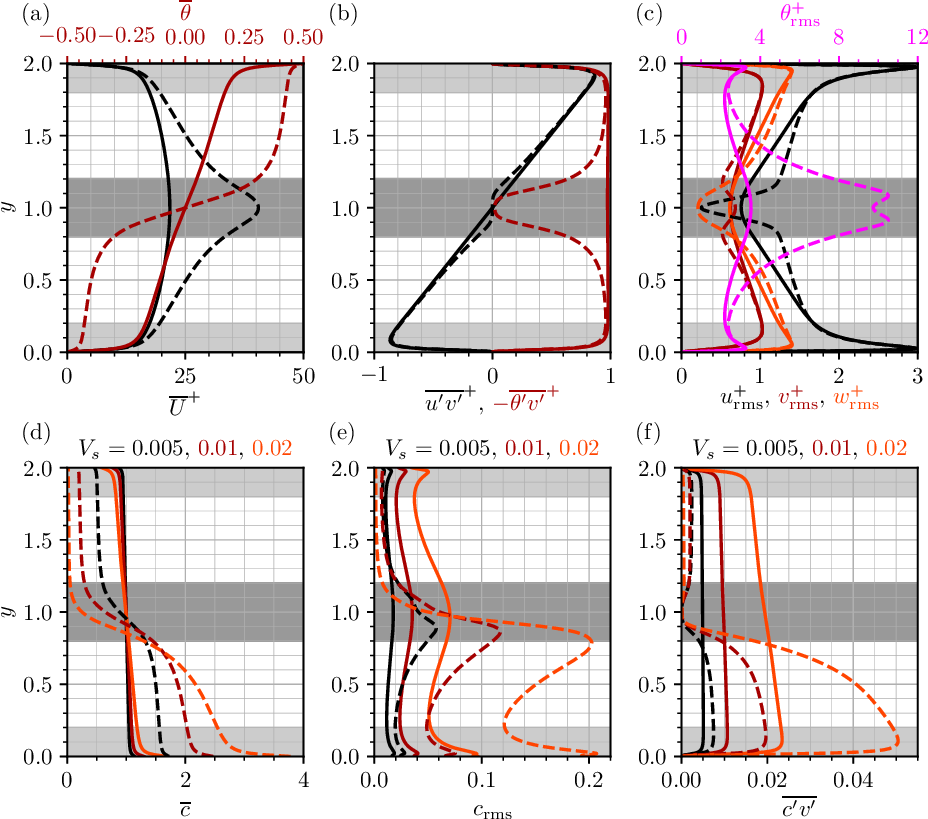}
    \caption{
        Time- and planar-averaged statistics for Case U (solid lines) and Case S (dashed lines).
        Colours represent different variables indicated by sub-figure titles, and the shaded regions represent the approximate bounds of the inner region ($y \leq 0.2$ and $y \geq 1.8$), the outer region ($0.2 < y \leq 0.8$ and $1.2 \leq y < 1.8$), and the channel core ($0.8 > y > 1.2$).
    }
    \label{fig:profiles}
\end{figure}

Profiles of time- and planar-averaged statistics are presented in Figure \ref{fig:profiles}, where variables are decomposed into their time- and planar-averaged means (represented by overbars) and fluctuations away from the means (represented by primes): $U_i = \overline{U}_i + u_i'$, $\theta = \overline{\theta} + \theta'$, and $c = \overline{c} + c'$. 
Superscript $+$ represents normalisation in wall-units, where the appropriate scales for velocity and temperature are $u_\tau$ and $\theta_\tau$, where $\theta_\tau = \tfrac{1}{ \Rey_\tau \Prn} \left.\tfrac{ \partial \overline{\theta}}{ \partial y}\right|_w$.  
Case U is partitioned into two regions: The inner region ($y \leq 0.2$ and $y \geq 1.8$) where viscosity and shear are the key controls on dynamics, and the outer region ($0.2 < y < 1.8$) where the direct effects of viscosity are negligible.
This classification follows that typically used for boundary layer flows \citep[See e.g.][]{pope2001turbulent}.
Case S differs from Case U by introducing a further region, the channel core ($0.8 > y > 1.2$), where dynamics are dominated by buoyancy forces, indicated by the strong temperature gradients \citep{lloyd2022coupled}.

Before discussing the influence of density stratification on the sediment scalars, the key features of the stratified channel flow are summarised \citep{armenio2002investigation,garcia2011turbulence,lloyd2022coupled}.
When considering the temperature and velocity fields, key differences between unstratified and stratified cases occur in the outer and core regions.
Root-Mean-Square (RMS) temperature ($\theta_\text{rms}^2 = \overline{\theta'\theta'}$) and vertical velocity ($v_\text{rms}^2 = \overline{v'v'}$) fluctuations peak to local maxima in the channel core, while streamwise  ($u_\text{rms}^2 = \overline{u'u'}$) and spanwise  ($w_\text{rms}^2 = \overline{w'w'}$) velocity fluctuations are suppressed.
This is due to the internal waves that dominate the core, which suppress turbulence and vertical flux of momentum ($\overline{u'v'}^+$) and temperature ($\overline{\theta'v'}^+$).
The suppression of turbulence in the core leads to a jet-like peak in mean streamwise velocity, $\overline{U}^+$, approximately twice as high as the unstratified case.
The mean temperature $\overline{\theta}$ is approximately constant in much of the inner and outer regions for Case S, aside from in the very-near-wall region.
Strong temperature gradients are present in the channel core, responsible for the suppression of turbulence and vertical transport of momentum and temperature, and for enabling the propagation of internal waves.

Sediment concentration is approximately linear across most of the channel for Case U, aside from steep gradients near the wall.
Local peaks in $c_\text{rms}$ are observed near the walls and the near the channel centreline.
The vertical sediment flux is also approximately linear for Case U, unsurprising when noting that the mean sediment concentration is governed by
\begin{equation}
\label{eq:verticalSedimentBalance}
\frac{1}{\Scn \Rey_\tau} \pdev{\overline{c}}{y} + V_s \overline{c} - \overline{c'v'} = 0.    
\end{equation}
For Case U diffusive processes are negligible except near the walls (where sediment concentration gradients are large), such that $\overline{c'v'} \approx V_s \overline{c}$. 
Increasing $V_s$ steepens the sediment concentration gradient, increasing the sediment concentration at the lower wall and reducing it at the upper wall.

\begin{figure}
    \centering
    \includegraphics{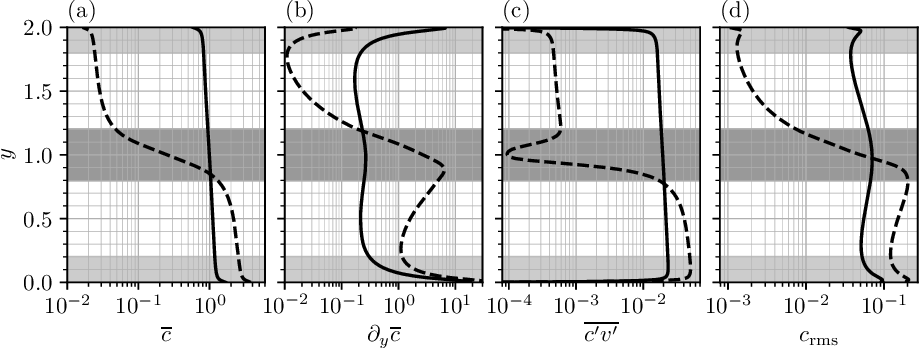}
    \caption{Time- and planar-averaged sediment statistics for Case U (solid lines) and Case S (dashed lines) with $V_s = 0.02$, logarithmically scaled.}
    \label{fig:profiles_sediment_log}
\end{figure}

As hypothesised by \citet{dorrell2019self}, stratification has a profound impact on sediment transport. 
The region of strong buoyancy gradient acts as a transport barrier, leading to a two-layer sediment concentration profile with high concentration in the lower region of the channel, and very low sediment concentrations in the upper regions of the channel. 
This is better visualised in Figure \ref{fig:profiles_sediment_log} where the sediment statistics (for $V_s = 0.02$) are scaled logarithmically. 
Here, the sediment concentration varies by two orders of magnitude between the upper and lower regions of the channel.
These regions are separated by a strong vertical concentration gradient, with a local maximum value near the lower edge of the channel core, coinciding with a peak in $c_\text{rms}$.
The sediment vertical flux is unsurprisingly similar in profile to the sediment concentration, but has a local minimum at the channel centreline. 
$V_s$ appears to have a much stronger influence on sediment statistics than for unstratified flow (Figure \ref{fig:profiles}); at higher $V_s$ there is a much larger difference between sediment concentrations below/above the channel core, and much larger fluctuations in sediment concentrations. 
Increasing $V_s$ also shifts the local maxima for sediment concentration gradient and $c_\text{rms}$ downwards, further from the channel centreline. 

\begin{figure}
    \centering
    \includegraphics{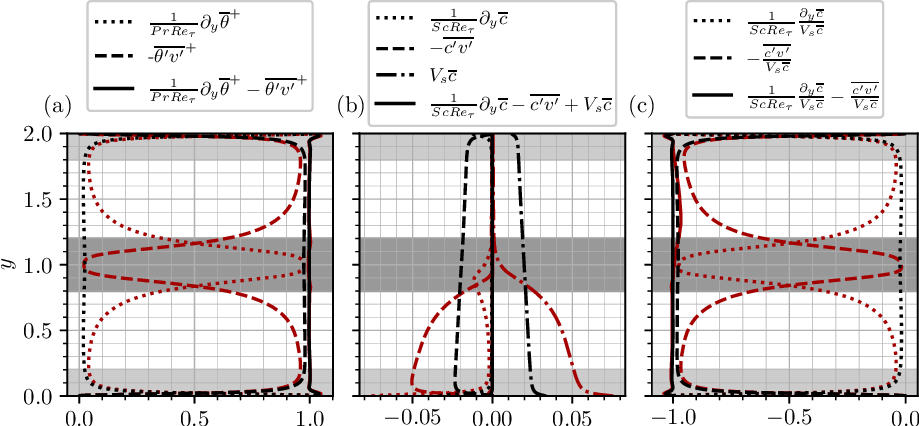}
    \caption{Budgets of transport equations for (a) temperature, (b) sediment with $V_s = 0.02$, and (c) sediment with $V_s=0.02$ and scaled by $V_s \overline{c}$. Transport terms correspond to different linestyles (see legends) and are as per equations \eqref{eq:verticalTemperaturePlusBalance} for (a), \eqref{eq:verticalSedimentBalance} for (b), and \eqref{eq:verticalSedimentPlusBalance} for (c). Colours represent Case U data (black) and Case S data (red).}
    \label{fig:budgets}
\end{figure}

A key difference between Case U and Case S scalar transport processes is the dominance of `diffusive' processes, as opposed to turbulent mixing processes, in the quasi-laminar core of the stratified flow. 
This is evident when assessing the vertical transport budgets of sediment concentration \eqref{eq:verticalSedimentBalance} and temperature: 
\begin{equation}
\label{eq:verticalTemperatureBalance}
\frac{1}{\Prn \Rey_\tau} \pdev{\overline{\theta}}{y} - \overline{\theta'v'} = \text{const.} = \theta_\tau,  
\end{equation}
or alternatively 
\begin{equation}
\label{eq:verticalTemperaturePlusBalance}
\frac{1}{\Prn \Rey_\tau} \pdev{\overline{\theta}}{y}^+ - \overline{\theta'v'}^+ = 1.  
\end{equation}
Budgets of vertical sediment concentration \eqref{eq:verticalSedimentBalance} and temperature \eqref{eq:verticalTemperatureBalance} transport are presented in Figure \ref{fig:budgets}.
The terms $\overline{c'v'}$ and $\overline{\theta'v'}$ are suppressed in the channel core for Case S, where the vertical transport of both sediment and temperature is governed by respective mean vertical gradients. 
Sediment and temperature scalar transport equations differ, clearly, due to the settling term, $V_s \overline{c}$. 
Interestingly, when scaling the vertical transport equation for sediment concentration by $V_s \overline{c}$ one obtains essentially the same balance between turbulent and diffusive processes as the temperature transport equation throughout the full channel height, aside from a change in sign (due to the increase in temperature with $y$, as opposed to the decrease in sediment concentration) as observed in Figure \ref{fig:budgets}.
When written explicitly we see
\begin{equation}
\label{eq:verticalSedimentPlusBalance}
\frac{1}{V_s \overline{c}}\frac{1}{\Scn \Rey_\tau} \pdev{\overline{c}}{y} - \frac{1}{V_s \overline{c}}\overline{c'v'} = -1    
\end{equation}
which clearly has similarities to \eqref{eq:verticalTemperaturePlusBalance}, which balances diffusive and turbulent components.
Functionally these equations are similar, given that the sediment scalars are dictated by the velocity field which is coupled to the temperature field. 
Further, the symmetry about the channel centreline for \eqref{eq:verticalSedimentPlusBalance} budgets is well explained when noting that at both walls, we must have $\tfrac{1}{{ V_s \overline{c}}}\tfrac{1}{{ \Scn \Rey_\tau}} \partial_y\overline{c} \approx -1 $, and Case U must have $- \tfrac{1}{{ V_s \overline{c}}}\overline{c'v'} \approx -1$ everywhere else.
Case S differs from Case U in the outer and core regions of the channel, particularly in the core where diffusive processes dominate with $\tfrac{1}{{ V_s \overline{c}}}\tfrac{1}{{ \Scn \Rey_\tau}} \partial_y\overline{c} \approx -1 $.
Under the scalings of $\theta_\tau$ and $V_s \overline{c}$ there are obvious similarities between scalar transport processes in the cases where either turbulent or diffusive components dominate.
For example, when turbulent processes dominate $\overline{\theta'v'}^+ \approx - \tfrac{1}{{ V_s \overline{c}}}\overline{c'v'} \approx -1$.
It is less obvious how scalars should compare when both of the two contributions are relevant.

It should be noted that the budgets do not balance perfectly in Figure \ref{fig:budgets}, primarily near the walls. This is due to the unresolved `sub-grid-scale' energy that leads to an underprediction of near-wall TKE \citep{lloyd2022coupled}. 
There is, however, an additional source of discrepancy for Case S that is revealed when scaling budgets according to \eqref{eq:verticalSedimentPlusBalance}, which shows that in the upper region of the channel the pseudo steady-state vertical transport equation is not entirely satisfied (up to a \SI{3}{\%} error). 
This is due to the time-scales required to obtain statistically steady solutions for the sediment scalars, which require $\mathcal{O}(\delta/V_s)$ timescales to converge.
This is further complicated by the transport barrier in the channel core which restricts the vertical sediment flux, therefore slowing convergence further.
Finally, we note that the sediment concentration is two orders of magnitude lower in the upper region of the channel than the lower region, and therefore the scaling $1/V_s \overline{c}$ becomes very large, thus enhancing any discrepancies in transport budgets.
Balances would be better satisfied if simulations were allowed to evolve further, but statistics are well converged outside this region (and indeed the regions where sediment concentration is non-negligible), and are therefore adequate for the present study.

\begin{figure}
    \centering
    \includegraphics{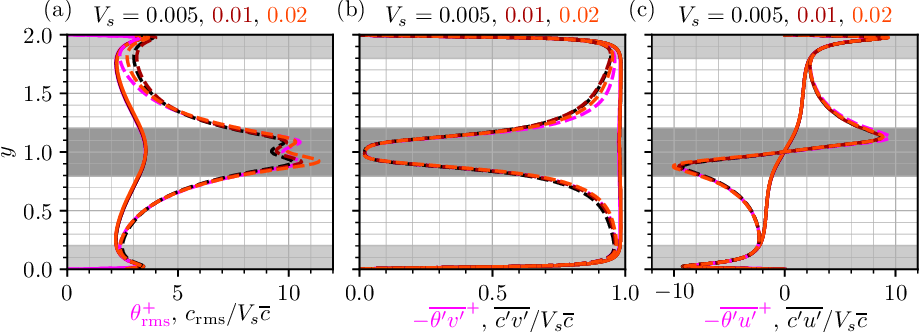}
    \caption{Temperature and sediment concentration statistics, with temperature statistics scaled by $\theta_\tau$ and concentration statistics scaled by $V_s \overline{c}$. Solid lines represent Case U data, dashed lines represent Case S data.}
    \label{fig:sediment_scaled}
\end{figure}

The scalings $\theta_\tau$ and $V_s \overline{c}$ appear to collapse statistics for $\theta_\text{rms}$, $\overline{\theta'u'}$, and $\overline{\theta'v'}$, and equivalent sediment concentration statistics, as shown in Figure \ref{fig:sediment_scaled}. 
Aside from some slight deviations in the channel core (particularly at the maxima of $\theta_\text{rms}$) and the upper region for Case S where $\overline{c}$ is small, the collapse between different scalars is remarkable.
Note that the deviations in statistics for $y > 1.2$ are most likely due to the convergence issues discussed, particularly when noting that the sediment field statistics tend closer to the temperature field statistics as $V_s$ increases. One would expect discrepancies between statistics to increase as $V_s$ increases, given this is parameter that controls the difference between temperature and sediment transport, but this result can be justified by noting that as $V_s$ increases the timescales required for convergence ($\delta/V_s$) decrease, and therefore cases with higher $V_s$ are better converged, and in better agreement with the temperature field statistics.
The collapse of the appropriately scaled vertical scalar fluxes can be explained by the balance of respective scalar transport equations, \eqref{eq:verticalTemperaturePlusBalance} and \eqref{eq:verticalSedimentPlusBalance}, and previous discussions. 
What is less clear is why the other statistics appear to collapse, aside from deviations in the channel core.

\begin{figure}
    \centering
    \includegraphics{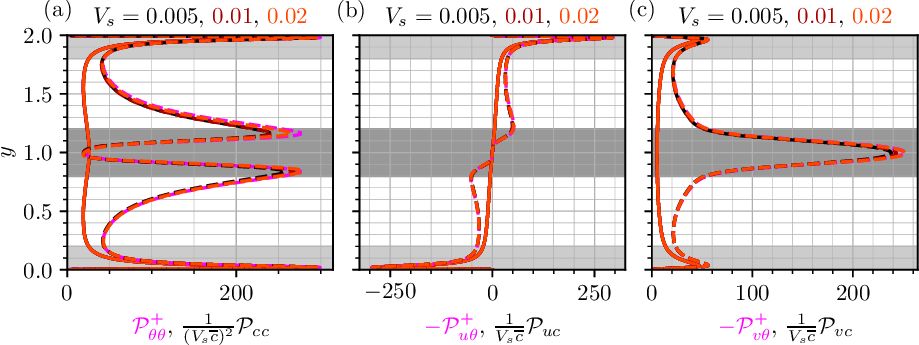}
    \caption{Production of passive (sediment concentration) and active (temperature) scalar fluxes and fluctuations, scaled by $\theta_\tau$ for temperature and $V_s \overline{c}$ for sediment. Solid lines represent Case U data, dashed lines represent Case S data.}
    \label{fig:production}
\end{figure}

\begin{figure}
    \centering
    \includegraphics{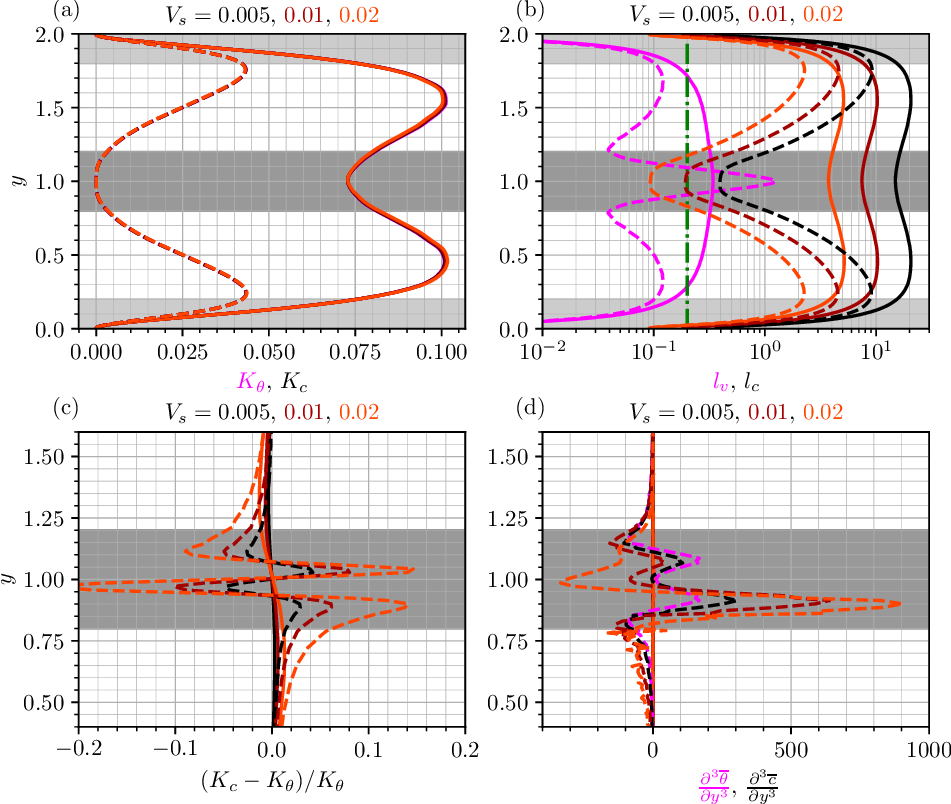}
    \caption{Temperature and sediment concentration (turbulent) diffusivities (a), and relative differences between them (c). Panel (b) estimates vertical turbulent mixing length scales ($l_v$) and sediment gradient length scales ($l_c$). Solid lines represent Case U data and dashed lines represent Case S data. The vertical dashed line of Panel (c) marks a length scale of $l_v=0.2$, equal to the half-width of the channel core.
    Panel (d) plots the third-order vertical gradients of mean temperature and sediment concentration.
    }
    \label{fig:diffusivities}
\end{figure}

Some insight can be gained by assessing the production terms for the transport of each of these statistics:
\begin{equation}
    \mathcal{P}_{c u} = - \overline{c' v'} \pdev{\overline{U}}{y} - \overline{u' v'}\pdev{\overline{c}}{y} , \ \ \ 
    \mathcal{P}_{c v} = - \overline{v' v'}\pdev{\overline{c}}{y} , \ \ \ 
    \mathcal{P}_{c c} = - 2\overline{c' v'} \pdev{\overline{c}}{y},
\end{equation}
with equivalent production terms for $\theta$ statistics obtained by substituting $c'$ and $\overline{c}$ for $\theta'$ and $\overline{\theta}$.
The collapse of statistics in Figure \ref{fig:sediment_scaled} implies that production terms of respective statistics should also scale similarly, such that
\begin{equation}
    \mathcal{P}_{\theta u}^+ \approx -\frac{1}{V_s \overline{c}}\mathcal{P}_{cu}, \ \ \ 
    \mathcal{P}_{\theta v}^+ \approx -\frac{1}{V_s \overline{c}}\mathcal{P}_{cv}, \ \ \ 
    \mathcal{P}_{\theta \theta}^+ \approx \frac{1}{(V_s \overline{c})^2}\mathcal{P}_{cc},
\end{equation}
noting that $c_\text{rms}^2 = \overline{c'c'}$ and $\theta_\text{rms}^2 = \overline{\theta'\theta'}$. 
Indeed, Figure \ref{fig:production} demonstrates that the production terms do collapse subject to this scaling, aside from some small deviations in the channel core.
The scalar gradients and fluxes are intricately connected and scale in the same way, readily explained by the gradient diffusion hypothesis.
Following \citet{nielsen2004turbulent}, a Taylor series expansion of a concentration field at $y \pm l_m/2$, with $l_m$ denoting a mixing length, in a flow with no background gradients in vertical velocity, leads to a vertical sediment flux
\begin{equation}
    \label{eq:taylor}
    c'v' = v'\left(c[z-l_m/2] - c[z+l_m/2]\right) = v'\left(l_m \pdev{c}{z} + \frac{l_m^3}{24} \frac{\partial^3 c}{\partial z^3} + \dots\right).
\end{equation}
To leading order, 
\begin{equation}
    c'v' = l_m v' \pdev{c}{z} = K_c \pdev{c}{z},
\end{equation}
where $K_c$ represents turbulent diffusion.
The vertical sediment (and indeed temperature) flux is therefore intricately linked to the vertical sediment concentration gradient, if gradient diffusion holds.
We should therefore expect to see the mean concentration gradient scale like the mean sediment flux, assuming that $K_c$ has no dependence on concentration or $V_s$.
In fact, we observe that $K_c$ and $K_\theta$ (equivalent turbulent diffusion for temperature) collapse to approximately the same curve for all scalars, for both Case U and Case S, to leading order (Figure \ref{fig:diffusivities}). 
Here we have used the definitions $K_c = -\overline{c'v'}/\partial_y \overline{c}$, and  $K_\theta = -\overline{\theta'v'}/\partial_y \overline{\theta}$.
The leading order collapse of all scalar turbulent diffusivities implies that the mean scalar gradients are strongly related to the mean vertical sediment flux, such that $\partial_y \overline{\theta}^+ \approx \partial_y \overline{c} / V_s \overline{c}$, given that $\overline{\theta'v'}^+ \approx \overline{c'v'}/V_s \overline{c}$.
The collapse of the scalar turbulent diffusivities explains the close agreement between their respective turbulent statistics, when appropriately scaled (Figures \ref{fig:sediment_scaled} and \ref{fig:production}). 

However, deviations between scalar transport statistics are present in the core of the channel, particularly for $c_\text{rms}/V_s \overline{c}$ and $\theta_\text{rms}^+$ at $y \approx 0.9$, and for $\overline{c'u'}/V_s \overline{c}$ and $\overline{\theta'u'}^+$ at respective maxima/minima. 
Note that all statistics collapse to single profiles for unstratified flow; strong buoyancy gradients are the most likley cause for the separation between the different scalars for Case S. 
A possible cause for this can be revealed by noting that despite seeming indistinguishable, the $K_c$ and $K_\theta$ profiles actually harbour large differences in the channel core, particularly in the region where both diffusive and turbulent components of \eqref{eq:verticalTemperaturePlusBalance} and \eqref{eq:verticalSedimentPlusBalance} are important (at $y \approx 0.9$, turbulent fluxes contribute approximately \SI{20}{\%} to the total budget, shown in Figure \ref{fig:budgets}). This is shown explicitly in Figure \ref{fig:diffusivities} (b) by plotting the relative difference between scalar diffusivities, $(K_c - K_\theta) / K_\theta$. 
Differences increase with increasing $V_s$, reaching \SI{20}{\%} for $V_s = 0.02$ when stratification is present.
Without strong buoyancy gradients (Case U), differences in scalar turbulent diffusivities are much smaller.
We hypothesise that these differences arise due to a violation of the gradient diffusion hypothesis that appears vital for the scalings by $V_s \overline{c}$ and $\theta_\tau$ to collapse. 
In particular, the higher order terms of the Taylor series expansion \eqref{eq:taylor} can only be neglected if the mixing length $l_m$ is much smaller than the indicative length scale over which the concentration field varies.
When stratification is present, and at higher values of $V_s$ the concentration field varies greatly over small distances at the edge of the channel core, yet the length scales over which turbulent mixing occurs are relatively large. We shall provide evidence for this over the subsequent sections, although a first approximation of these scales are shown in Figure \ref{fig:diffusivities} (c), where $l_c = (\partial_y \overline{c})^{-1} / \overline{c}$ represents the length scale over which the mean sediment concentration varies and $l_v = v_\text{rms}^{3/2}/\varepsilon$ represents the vertical turbulent mixing scale, where $\varepsilon$ is the turbulent kinetic energy dissipation rate. 
For illustrative purposes, we have further included a line at $l_v = 0.2$, equal to the half-width of the core, as a crude estimate of the mixing lengths that may be expected in that region. 
$l_c$ is very large across much of the channel, particularly for Case U scalars, indicating very slow variation in concentration.
In addition, $l_c >> l_v$ for all Case U scalars, such that higher order terms in the Taylor series expansion \eqref{eq:taylor} can be neglected, and gradient diffusion appears valid.
Both Case S and Case U have $l_c$ decreasing for increasing $V_s$, consistent with higher concentration gradients.
Finally note that for Case S, we have $l_v > l_c$ in the channel core.
For $V_s = 0.02$ we find $l_c$ is approximately half the value of $l_v$ at $y \approx 0.9$, indicating that vertical mixing scales are indeed large compared to the sediment concentration length scales. 
Of course, at the centre of the channel turbulent mixing processes have a negligible influence on dynamics due to their suppression by buoyancy forces, but closer to the edges of the core (e.g. $y \approx 0.9$) turbulence has a leading order effect.
The large-scale mixing processes, relative to the concentration gradients, may therefore explain the differences observed between the passive scalar statistics.
Further evidence of the importance of higher order terms in the calculation of vertical diffusivities are the third order time- and planar-averaged scalar gradients of Figure \ref{fig:diffusivities} (d). 
For Case S there are significant deviations between the different scalars, with differences increasing as $V_s$ increases.
Further, differences appear well correlated with differences in $K_\theta$ and $K_c$. 
These results therefore suggest that the coincidence of large-scale mixing events and strong concentration gradients leads to a deviation from first order gradient diffusion.

\subsection{Spectra}
\label{section:spectra}
\begin{figure}
    \centering
    \includegraphics{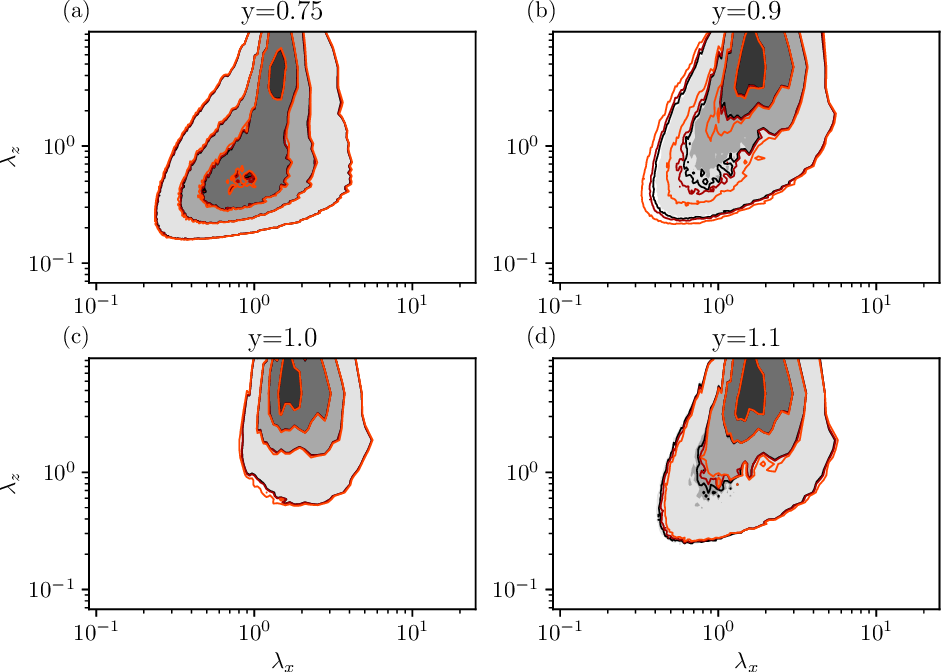}
    \caption{Premultiplied 2D energy spectra of scalar fluctuations as a function of streamwise and spanwise wavelengths at four $y$ positions. Grey filled contours represent temperature spectra, $k_x k_z E^{2D}_{\theta \theta}$, and coloured contour lines represent sediment concentration spectra, $k_x k_z E^{2D}_{cc}$, for $V_s = 0.005$ (black), $V_s = 0.01$ (red), and $V_s = 0.02$ (orange). Contour values represent \SI{20}{\%}, \SI{40}{\%}, \SI{60}{\%}, and \SI{80}{\%} of respective maxima.}
    \label{fig:2D_spectra}
\end{figure}

2D energy spectra for the scalar fluctuations are presented in Figure \ref{fig:2D_spectra}.
Here we calculate energy spectra as a function of streamwise and spanwise wavelength, $\lambda_x$ and $\lambda_z$, on several $y-$normal slices.
The contours presented represent \SI{20}{\%}, \SI{40}{\%}, \SI{60}{\%}, and \SI{80}{\%} of respective maxima. 
First note that we only report spectra in and near the channel core, finding that all scalars collapse to the same spectra in the majority of the channel.
An example of the clear collapse of all scalars is in panel (a) of Figure \ref{fig:2D_spectra}, where all four scalars show two distinct peaks in spectral energy.
The higher $\lambda_z$ peak is associated with the large-scale internal waves in the channel core, while the lower $\lambda_z$ peak is associated with turbulent activity \citep{garcia2011turbulence,lloyd2022coupled}.
The peak associated with turbulence subsides in the channel core, until only the peak associated with internal waves is present at $y=1$.
At $y=1$ we see all scalars collapse onto essentially the same energy spectra contours, demonstrating that the effect of internal waves is essentially the same for all scalar transport, relative to respective spectral maxima. 
However, deviations between the scalars are clear closer to the edges of the core, at $y=0.9$ and $y=1.1$. 
While the peak associated with the internal waves shows reasonable collapse between the scalars, the `tail' of the spectra deviates significantly, with larger differences associated with larger $V_s$. 
This indicates that it is the turbulent structures that are the cause of deviations between scalar fluctuations, not the largely reversible internal wave motions.
Deviations of $E_{cc}^{2D}$ from $E_{\theta \theta}^{2D}$ manifest in a broadening spectral peak for $y=0.9$ as $V_s$ increases, and a sharpening spectral peak at $y=1.1$. In other words, the contour lines shift outwards as $V_s$ increases at $y=0.9$, but inwards at $y=1.1$. 

\begin{figure}
    \centering
    \includegraphics{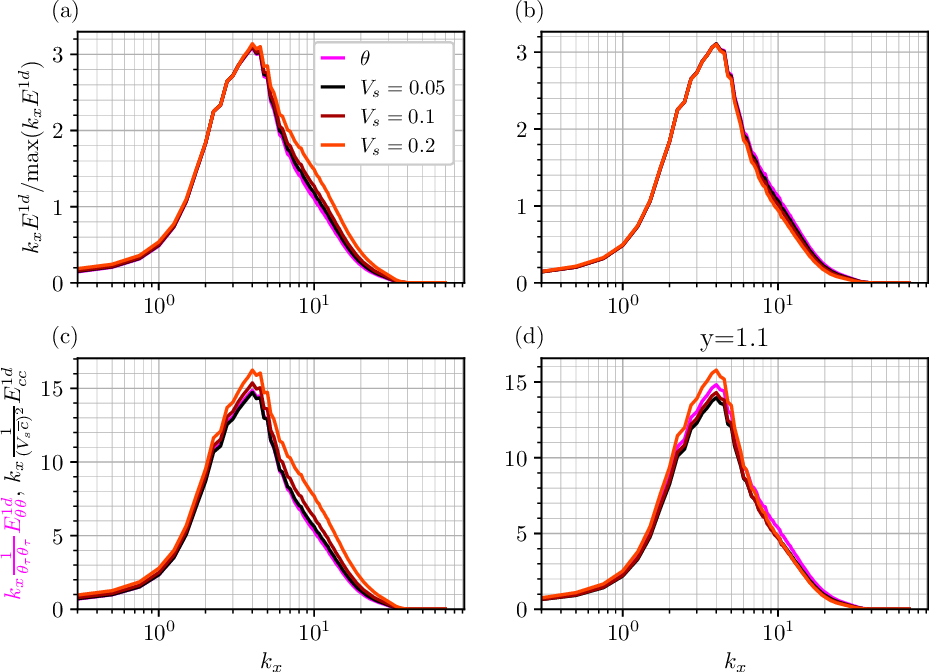}
    \caption{Premultiplied 1D energy spectra of scalar fluctuations at $y=0.9$ in (a) and (c), and $y=1.1$ in (b) and (d), as a function of streamwise wavenumber. Spectra are normalised by respective maxima in panels (a) and (b), and $\theta_\tau^2$ for temperature spectra and $\left( V_s \overline{c}\right)^2$ for sediment concentration spectra in (c) and (d). Lines are as per the legend in panel (a).}
    \label{fig:1D_spectra}
\end{figure}

1D Spectra at $y=0.9$ and $y=1.1$ are reported in Figure \ref{fig:1D_spectra}, as a function of streamwise wavenumber $k_x = 2\pi/\lambda_x$. 
In addition to normalisation by respective maxima, spectra are shown scaled by $\theta_\tau$ for temperature and $V_s \overline{c}$ for sediment concentration.
Both scalings lead to qualitatively similar behaviour.
At $y=0.9$ higher $V_s$ leads to a broader range of energetic scales, particularly at the higher wavenumbers associated with turbulence. 
At $y=1.1$ higher $V_s$ leads to a sharper peak associated with the internal waves, with a reduced spectral energy associated with turbulence.
We speculate that this difference arises due to the change in sign of $\partial^3_y \overline{c}$ (Figure \ref{fig:diffusivities}) above and below the channel centreline, particularly for $V_s = 0.02$.
Indeed, we observe larger differences in scalar transport spectra at $y=0.9$ than at $y=1.1$, agreeing well with the differences in $\partial^3_y \overline{c}$ and $\partial^3_y \overline{\theta}$ at those locations.
This therefore supports the argument that it is a breakdown of the leading order gradient diffusion hypothesis that causes deviations between scalar transport statistics.

\begin{figure}
    \centering
    \includegraphics{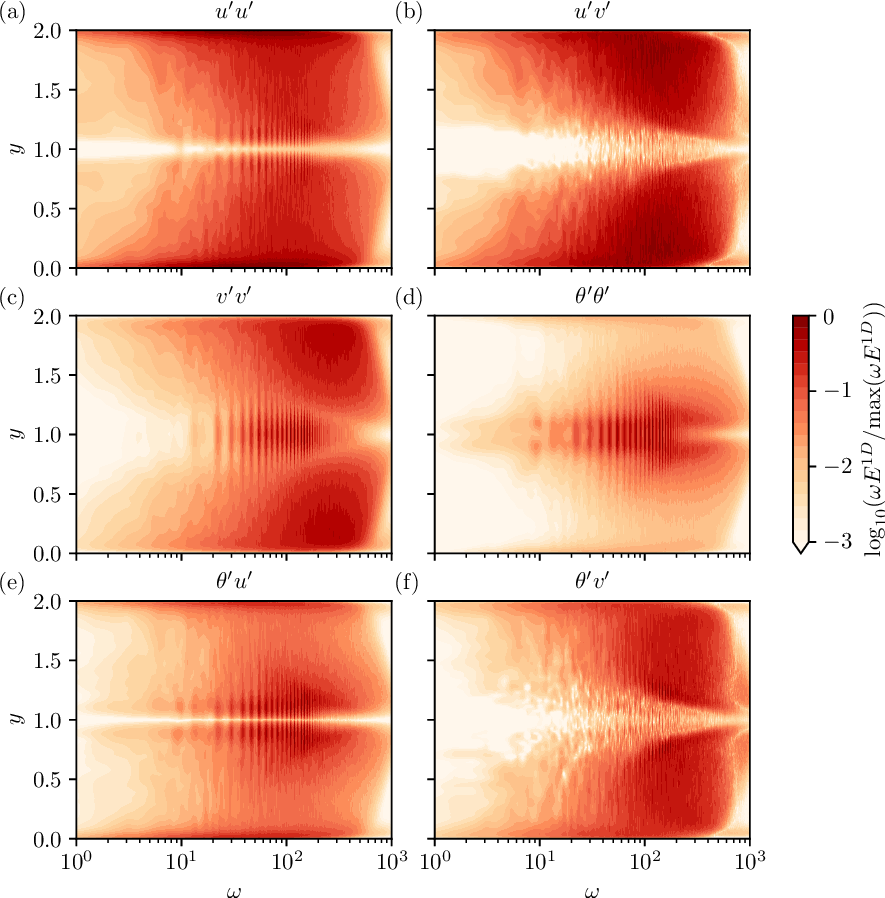}
    \caption{Premultiplied 1D energy spectral magnitude as a function of temporal frequency and vertical coordinate for Case S. Panels (a) to (f) represent different statistics for fluid velocity, temperature, and associated fluxes. All spectra are normalised by respecive maxima.}
    \label{fig:temporal_spectra_temp}
\end{figure}

\begin{figure}
    \centering
    \includegraphics{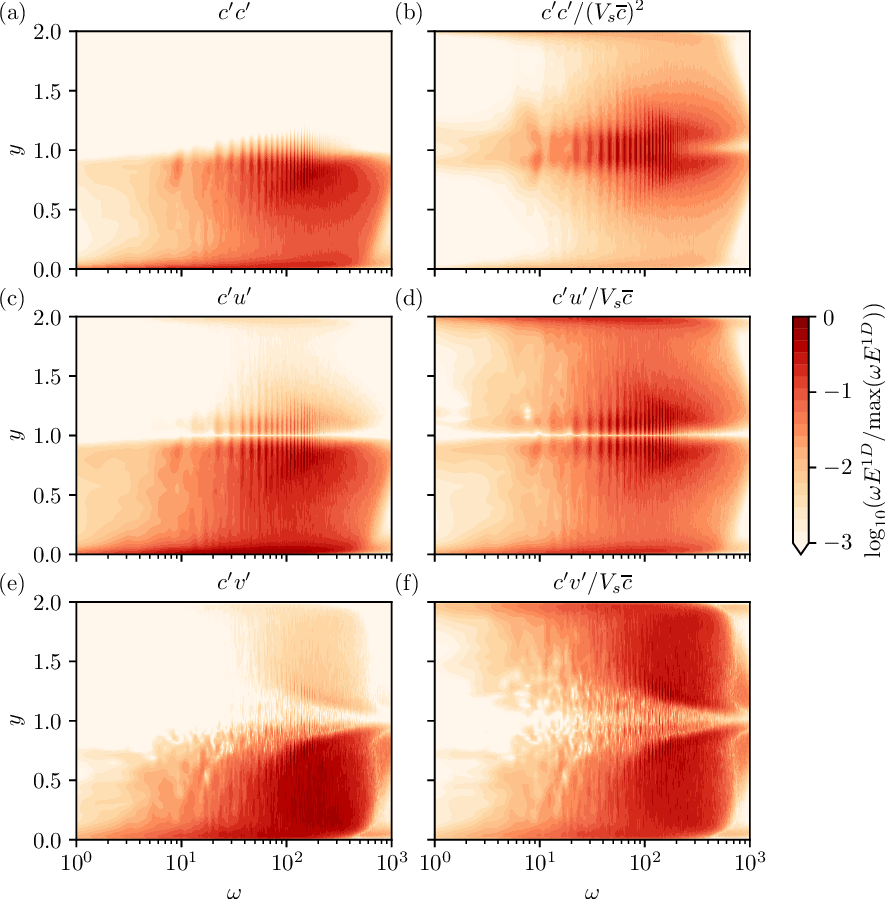}
    \caption{Premultiplied 1D energy spectral magnitude as a function of temporal frequency and vertical coordinate for Case S. Panels (a), (c), and (e) present spectra associated with sediment concentration fluctuations, streamwise concentration flux, and vertical concentration flux, respectively, for $V_s = 0.02$. Panels (b), (d), and (f) present the same spectra as panels (a), (c) and (e) but scaled by $V_s \overline{c}$. All spectra are normalised by respective maxima.}
    \label{fig:temporal_spectra_conc}
\end{figure}

1D Temporal spectra are presented in Figures \ref{fig:temporal_spectra_temp} and \ref{fig:temporal_spectra_conc}, presented as a function of vertical coordinate $y$ and frequency $\omega$, for Case S. 
Figure \ref{fig:temporal_spectra_temp} presents temperature and velocity statistics, clearly showing the extent and influence of the internal waves in the channel core.
Spectral energy for $u'u'$, and for streamwise and vertical momentum and temperature fluxes all show local minima at $y=1$. 
The streamwise flux of momentum and temperature is zero at $y=1$ for both Case S and Case U, owing to the symmetry of the flow and the local velocity maxima.
In contrast, the minima for vertical momentum and temperature flux, and streamwise velocity fluctuations arises due to the presence of strong buoyancy gradients and subsequent internal waves and suppression of turbulent irreversible motion.
Internal waves manifest as coherent and distinct peaks in spectral energy at discrete temporal frequencies, focussed in the channel core. 
The discrete temporal frequencies $\omega$ arise due to the finite spatial bounds of the periodic computational domain. Specifically, these peaks correspond to the convection of streamwise wavenumbers $k_x$ in multiples of $2\pi/L_x$ \citep{lloyd2022coupled}. 
An interesting aspect of Figure \ref{fig:temporal_spectra_temp} is the clear vertical bounds of the distinct and discrete peaks associated with internal waves, complementing the spectral and dynamic mode decomposition data of \citet{lloyd2022coupled}.
Figure \ref{fig:temporal_spectra_temp} also shows that buoyancy gradients in the channel core suppress the large-scale momentum and vertical fluxes more than the small-scale (or high frequency) structures. 
The suppression of large-scale vertical fluxes arises due to the buoyancy and concentration fluctuations being out-of-phase with the vertical velocity fluctuations, due to the internal waves (this will be demonstrated using SPOD in Section \ref{section:SPOD}).

Temporal spectra for the sediment fluctuations and fluxes are presented in Figure \ref{fig:temporal_spectra_conc}, for Case S and $V_s = 0.02$.
The imprint of the internal waves on sediment statistics is clear, with spectral amplitudes dramatically decreasing above the channel core, particularly for $c'c'$.
When scaled by $V_s \overline{c}$ sediment spectra are in clear agreement with the temperature statistics of Figure \ref{fig:temporal_spectra_temp}, further demonstrating the similarities between the transport of the different scalars.

\subsection{Spectral Proper Orthogonal Decomposition (SPOD)}
\label{section:SPOD}
To further establish the role internal waves have on sediment transport we adopt Spectral Proper Orthogonal Decomposition (SPOD) to identify the coherent motion in the channel core. 
SPOD is performed on $z-$ normal slice data using the method of \citet{schmidt2020guide}.
Data are organised into snapshots $\bm{q}(\bm{x},t)$ consisting of streamwise and vertical velocity, temperature, and sediment concentration for $V_s = 0.02$.
Following the method of Welch, snapshots are split into 35 consecutive and overlapping (with a \SI{50}{\%} overlap) blocks ($N_\text{blk} = 35$), each containing snapshots ($N_\text{fft} = 2048$).
Fourier transforms are subsequently performed on each block containing $N_\text{fft}$ snapshots, using a Hamming window, to obtain $\hat{\bm{q}}(\bm{x},\omega)$.

After obtaining $\hat{\bm{q}}$, analysis is performed one frequency at a time.
For each frequency the data are arranged into a matrix $\hat{Q}$,
\begin{equation}
    \hat{Q} = 
    \begin{bmatrix}
        \vdots & \vdots & & \vdots \\
        \hat{q}^1 & \hat{q}^2 & \dots & \hat{q}^N \\
        \vdots & \vdots & & \vdots 
    \end{bmatrix}, \ \ \hat{Q} \in \mathbf{C}^{M \times N}
\end{equation}
where $M$ is the total number of degrees of freedom and $N$ is the number of realisations (equal to the number of data blocks, $N_\text{blk}$) at the given frequency.
The SPOD modes, $\hat{\Phi}$, are eigenvectors of the sample cross-spectral density, given by
\begin{equation}
    \hat{C} = \frac{1}{N-1} \hat{Q} \hat{Q}^H,
\end{equation}
which are obtained by solving either
\begin{equation}
    \hat{C} W \hat{\Phi} = \hat{\Phi} \hat{\Lambda},
\end{equation}
or
\begin{equation}
    \hat{Q}^H W \hat{Q} \hat{\Psi} = \hat{\Psi} \hat{\Lambda}, \ \ \hat{\Phi} = \hat{Q} \hat{\Psi},
\end{equation}
where $W$ is a weighting matrix, and $\hat{\Lambda}$ represents the eigenvalues or energies associated with the SPOD modes $\hat{\Phi}$. 
In our case $M \gg N$ making the second formulation favourable, which computes the intermediate left singular vectors, $\hat{\Psi}$, before obtaining the SPOD modes. 
This approach avoids computation of the $M \times M$ cross-spectral density matrix $\hat{C}$ and the subsequently high-dimensional eigenvalue problem based on $\hat{\Phi}$.
The SPOD modes for a given frequency are orthonormal, with $\hat{\Phi}^H W \hat{\Phi} = I$. The weighting matrix $W$ is chosen to favourably identify coherent structures in the core of the channel, taking the value of 1 for $0.8 \leq y \leq 1.2$, and 0 elsewhere. Tests with and without this weighting function revealed that while the SPOD modes are qualitatively similar, low rank behaviour is more clear in the SPOD energies when modes are no longer polluted by near-wall data. 
For more information, see \citet{schmidt2020guide}.

\begin{figure}
    \centering
    \includegraphics{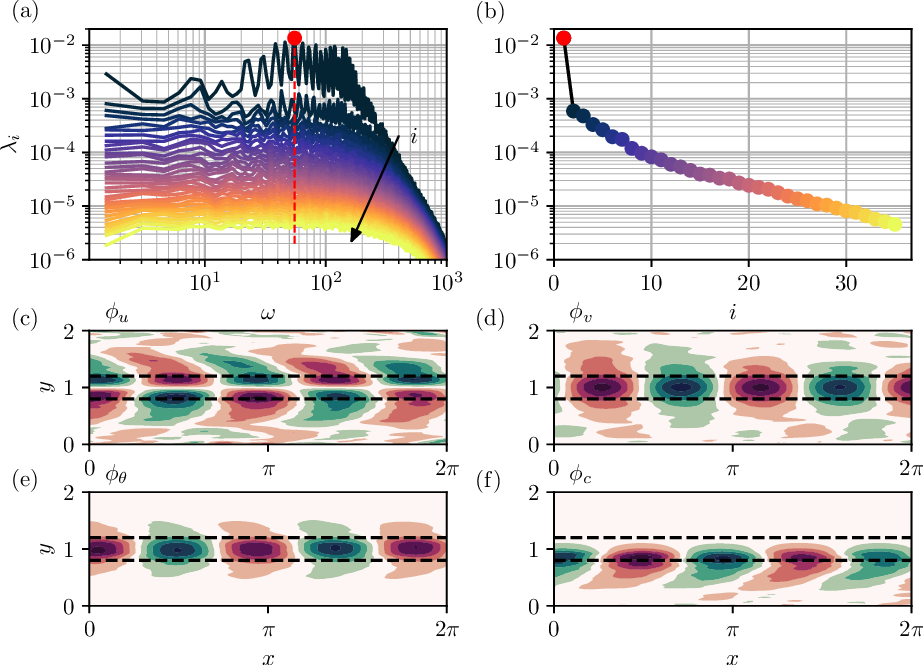}
    \caption{SPOD of Case S with $V_s = 0.02$, computed using $z-$normal slice data. Panel (a) reports modal energies $\hat{\Lambda}$ as a function of temporal frequency. 
    Lines and colours represent the mode number $i$, increasing from 1 to 35, with mode number 1 representing the modes with the highest energy for a given frequency.
    Modal energies at $\omega = 55.2$, indicated by the vertical dashed line in panel (a), are plotted against mode number in panel (b).
    The spatial structure of the highest energy mode at $\omega = 55.2$ (identified by the red marker in panels (a) and (b)) is reported in panels (c) to (f), where horizontal dashed lines represent the approximate bounds of the channel core region, $y=0.8$ and $y=1.2$. The streamwise velocity, vertical velocity, temperature, and sediment concentration components of $\hat{\Phi}$ are reported in panels (c) to (f), respectively.
    }
    \label{fig:spod}
\end{figure}

SPOD energies as a function of frequency can be observed in Figure \ref{fig:spod} (a).
For a given frequency SPOD modes are ranked based upon their energy, with mode 1 corresponding to the highest modal energy.
The lines of Figure \ref{fig:spod} (a) and markers of Figure \ref{fig:spod} (b) are coloured by the mode number $i$.
Distinct peaks in Mode 1 and Mode 2 are observed for $20 \lesssim \omega \lesssim 200$ corresponding to the internal waves in the channel core. 
These distinct peaks correspond to those observed in the temporal spectra of Figures \ref{fig:temporal_spectra_temp} and \ref{fig:temporal_spectra_conc}, and appear low-rank with mode 1 energies at least an order of magnitude higher than mode 2 energies.
Panel (b) of figure \ref{fig:spod} shows the modal energies associated with the SPOD for the frequency corresponding to the vertical dotted line in panel (a), where the low-dimensional nature of the SPOD is more clear. 
Interestingly, the peaks and troughs associated to the internal waves for mode 1 energies are reversed for mode 2 energies, with mode 2 energies typically decreasing when peaks in mode 1 are observed. This further demonstrates the low-rank nature of the internal waves in the channel core.
Panels (c) to (f) of Figure \ref{fig:spod} visualise the spatial structure of the SPOD mode corresponding to the modal energy identified by the red marker in panels (a) and (b). 
The $u$, $v$, $\theta$, and $c$ components of $\hat{\phi}$ are shown, in clear agreement with the dynamic mode decomposition data of \citet{lloyd2022coupled}.
The peaks and troughs in vertical velocity and temperature components are out of phase and concentrated in the core region $0.8 \lesssim y \lesssim 1.2$. The influence of the waves extends into the outer regions but the magnitude of the mode is significantly decreased.
The $c$ component of the SPOD modes is negatively correlated to the temperature field; peaks in $\Phi_\theta$ correspond to troughs in $\Phi_c$, owing to the difference in sign between respective wall-normal gradients.
The key difference between the different scalars is the downward offset of $\Phi_c$ peaks, which have their maximum value near the lower bound of the core, $y \approx 0.8$.
This offset is consistent with the local maxima of sediment RMS fluctuations and vertical sediment concentration gradient (Figure \ref{fig:profiles_sediment_log}). 
The `tails' of $\Phi_c$ extend further into the turbulent regions of the flow than $\Phi_\theta$, and peaks/troughs do not extend above the bounds of the core.
The infulence of channel core waves extends deeper into outer regions of the flow for high settling velocity sediments.
This effect is due to the large concentration gradient over the channel core; While not shown here, scaling the SPOD modes appropriately by $V_s \overline{c}$ unsurprisingly leads to a more symmetric mode, further demonstrating the importance of this scaling on concentration statistics.

SPOD clearly demonstrates that while the internal waves are prolific in the channel core, their influence on sediment statistics is not limited to that region, and that the coincidence of large-scale structures and steep concentration gradients may lead to non-trivial dynamics not captured by leading order gradient diffusion models. 

Note that the spatial structure for all internal wave modes are qualitatively similar and so are not shown here, differing only in the number of structures that fit in the domain (i.e their streamwise wavenumber $k_x$). 
Sensitivity of the decomposition to the parameters $N_\text{blk}$ and $N_\text{fft}$ was assessed by also computing SPOD for $N_\text{fft} = 4098$, reducing the number of blocks to $N_\text{blk} = 18$.
While results were qualitatively similar, some sensitivity was observed for the low frequency peaks in energy, where the higher $N_\text{fft}$ better resolved individual peaks at the expense of higher levels of noise.

\subsection{Instantaneous structure}
\label{section:inst}
\begin{figure}
    \centering
    \includegraphics{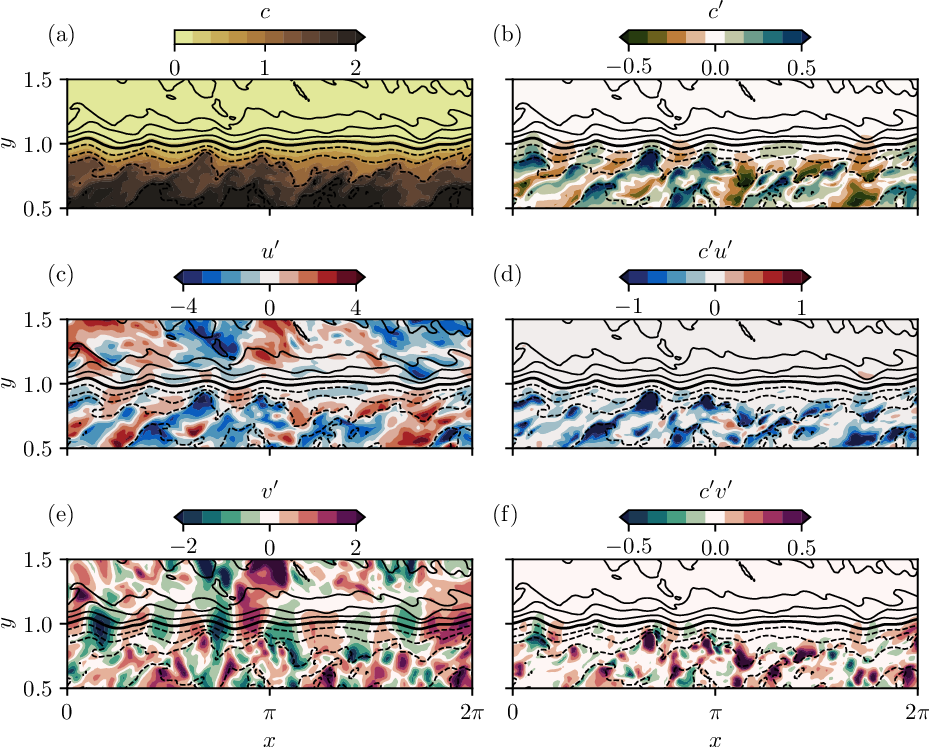}
    \caption{
    Instantaneous structure of Case S with $V_s = 0.02$ plotted using $z-$normal slice data. 
    Data are sediment concentration (a), sediment concentration fluctuations (b), streamwise velocity fluctuations (c), streamwise sediment flux (d), vertical velocity (e), and vertical sediment flux (f).
    Solid lines are temperature contours with negative values indicated by dashed lines, and the $\theta=0$ contour represented by a thicker line. 
    }
    \label{fig:inst}
\end{figure}

The instantaneous structure of Case S is shown in Figure \ref{fig:inst}, where a snapshot of the flow is visualised on a $z-$normal plane.
Temperature contours overlay all the figures with negative $\theta$ marked with dashed lines and positive $\theta$ with solid lines, clearly indicating the approximate bounds of the core region, where contours are approximately parallel and tightly packed (indicating the strong buoyancy gradient).
The internal wave field is most easily observed through the contours of $v'$, which show minima/maxima occurring throughout the channel core, approximately out of phase with the temperature contours. 
The general shape of the sediment concentration contours are in close agreement with the temperature field, particularly in the core where turbulence is suppressed, and wave elasticity can form a barrier to vertical scalar transport \citep{dorrell2019self}.
Of particular note is the clear downward offset of the $c'$ minima/maxima, in agreement with the statistics presented in the previous sections. 

The hairpin structures identified by \citet{lloyd2022coupled} can be identified in the sediment streamwise and vertical fluxes, which act to transport sediment into the channel core from the outer regions.
This is most clear from the $c'u'$ fluxes which indicate large peaks in negative $c'u'$, where relatively slow high concentration fluid has been lifted up from the outer portion of the flow. 
In the core $c'v'$ has local minima/maxima corresponding to these ejection events, although these upward/downward vertical fluxes are in approximate balance owing to the mostly reversible motions of the waves in the channel core. 
The vertical fluxes are however large in the outer regions of the flow and at the core edge, where ejections of hairpins contribute most to vertical sediment flux.

\section{Discussion and conclusions}
\label{section:conclusions}
We have simulated the passive transport of buoyant particles in a channel flow stratified by thermal gradients. 
Stratification has a profound impact on sediment transport in the core of the channel, where buoyancy forces dominate.
We find that stratification introduces a strong transport barrier in the core, leading to a two-layer sediment concentration profile with sediment trapped in the lower regions of the channel. 
Concentration gradients and fluctuations are enhanced by stratification, but are offset from the core centreline due to the enhanced concentration beneath the core, an effect that increases with the particle settling velocity, $V_s$.
In addition, the influence of $V_s$ on sediment statistics is enhanced by stratification due to the large differences in sediment concentrations above and below the channel core. 
SPOD and temporal spectra show that the influence of waves on sediment is offset downwards compared to temperature, due to the increased sediment concentration in the lower region of the channel; sediment fluctuations are negatively correlated to the temperature fluctuations, and as $V_s$ increases the peak in fluctuations due to waves shifts further downwards towards the core-edge, and the influence of waves is felt through more of the channel outer region. 

However, we find that the differences between scalar transport of the three sediments and the temperature field are approximately the same when appropriately scaled by $\theta_\tau$ for temperature and $V_s \overline{c}$ for the sediment concentrations. 
These scalings collapse all statistics to approximately the same vertical profiles and spectra. 
Given that only the temperature field is two-way coupled to the momentum equation, it follows that the sediments are indirectly coupled to temperature transport.
The approximate collapse of statistics for the different scalars can be explained by assessing their vertical transport budgets which have functionally the same form when appropriately scaled; a turbulent scalar flux balances a term proportional to the mean vertical gradient of the scalar.
In the case that either term dominates over the other it is clear that either $-\partial_y \overline{c} / \Scn V_s \overline{c} \approx \partial_y \overline{\theta}^+ / \Prn \approx 1$ or $\overline{v'c'} / V_s \overline{c} \approx -\overline{v'\theta'}^+ \approx 1$.
Further, turbulent statistics for the different scalars approximately collapse using the same scaling if the gradient diffusion hypothesis holds, given that production of turbulent fluctuations and fluxes are dependent on not only the scalar fluxes but also the scalar mean vertical gradients.

However, these scalings are unable to collapse statistics in and at the edges of the channel core.
Here, differences between scalars increase with increasing $V_s$; deviations in the scalar turbulent diffusivities reach \SI{20}{\%} when $V_s = 0.02$, in the channel core.
We argue that these deviations arise due to the importance of both molecular diffusive forces and turbulent fluxes, and the violation of the gradient diffusion hypothesis, where concentration length scales are small compared to mixing length scales. 
This is demonstrated by estimating length scales in Figure \ref{fig:diffusivities}, and inspection of instantaneous data in Figure \ref{fig:inst}, where large-scale coherent structures, ejected from the outer regions of the flow, are the main source of turbulent mixing in the core of the channel. 
In addition, deviations between $\partial^3_y \overline{\theta}$ and $\partial^3_y \overline{c}$ become very large in the channel core, which are the first neglected derivatives in the Taylor Series expansion for $\overline{\theta'v'}$ and $\overline{c'v'}$ if the leading order gradient diffusion hypothesis were to hold. 
Further, scalar statistics separate at $y  \approx 0.9$ and $y \approx 1.1$ (Figure \ref{fig:sediment_scaled});
When inspecting spatial spectra at these locations we see that differences between the scalars arise due to turbulent scales rather than at the wavenumbers associated with internal waves, supporting the conclusion that it is turbulent mixing scales that lead to the differences between scalar transport. 
We therefore argue that it is the breakdown of the gradient diffusion hypothesis in the channel core, or the importance of higher order terms in the Taylor series expansion \eqref{eq:taylor}, that is the primary cause of the separation between the different scalars.

To adequately model flows with strong concentration gradients that coincide with large-scale mixing events a higher order diffusivity coefficient is required. 
While this study presents an idealised sediment transport problem the dynamics explored are commonplace in natural and industrial flows. 
Stratification, whether imposed through background variations in composition or suspended sediments, dominate dynamics in regions of strong buoyancy gradients, leading to scalar transport barriers where concentration gradients are large \citep{dorrell2019self}.
Relatively large-scale mixing events will therefore lead to deviations in vertical scalar turbulent diffusivities, and leading order gradient diffusion models will be unable to accurately predict dynamics \citep{nielsen2004turbulent}.
This study was limited to small $V_s$ but we demonstrate that increasing $V_s$ leads to larger concentration gradients and therefore smaller concentration length scales compared to turbulent mixing length scales.
We therefore speculate that further increase of $V_s$, through an increase of the particle size or density, will have a larger control on sediment transport.
We further speculate that $\Scn$ will have a key control on dynamics in the core region, given the dominance of (molecular) diffusive processes.
While it is often argued that $\Scn$ has a small effect on transport dynamics \citep[See e.g.][]{necker2005mixing, nasr2011turbins,nasr2014turbidity, pelmard2018grid}, it will have a larger influence on regions where turbulence is suppressed \citep{marshall2021effect}.
Like $V_s$, an increase in $\Scn$ will sharpen sediment concentration gradients in the channel core, and therefore likely cause further deviations between scalar transport and leading order gradient diffusion models.
Future work should also explore the effects of higher concentration particulates that feed-back on the fluid momentum via an additional buoyancy forcing term. In this way, stratification could be imposed by the sediment concentration field \citep{cantero2009direct}, or a double-diffusive problem could be formulated.

\section{Acknowledgements}
CJL was supported by an Early Career Fellowship funded by the Leverhulme Trust.
RMD would like to acknowledge funding support from NERC Independent Research Fellow Grant NE/S014535/1.
We acknowledge the Viper High Performance Computing facility of the University of Hull and its support team.

\bibliography{literature.bib}

\end{document}